\documentstyle[twoside,fleqn,npb,epsfig]{article}
\newdimen\picraise
\newcommand\picbox[1]
{
  \setbox0=\hbox{\input{#1}}
  \picraise=-0.5\ht0 
  \advance\picraise by 0.5\dp0
  \advance\picraise by 3pt      % correct for height of `=' above baseline
  \hbox{\raise\picraise \box0}
}
\def\kf{{\bf k}}
\newcommand{\AmS}{{\protect\the\textfont2
  A\kern-.1667em\lower.5ex\hbox{M}\kern-.125emS}}

\title{
    \begin{flushright}
      \normalsize
      \vspace{-1.4cm}
      Cavendish--HEP--99/04\\
      DAMTP--1999--65\\
      hep-ph/9905575
      \vspace{0.5cm}
    \end{flushright}
Elements of a field theory of unitarity corrections\thanks{Talk 
presented at the $7$th International Workshop on Deep Inelastic Scattering, 
Zeuthen, Germany, April 1999.}
}
\author{Carlo Ewerz\,\address{Cavendish Laboratory, Cambridge University, 
             Madingley Road, Cambridge CB3 0HE, U.\,K. and DAMTP, Cambridge University,
             Silver Street, Cambridge CB3 9EW, U.\,K.}%
        \thanks{Work supported in part by the EU Fourth Framework Programme
          `Training and Mobility of Researchers', Network `Quantum Chromodynamics
          and the Deep Structure of Elementary Particles',
          contract FMRX-CT98-0194 (DG 12 - MIHT).}
}

\begin{document}
%%%%%%%% take out for proc
\thispagestyle{empty}
%%%%%%%%

\begin{abstract}
I explain the motivation for investigating the high energy limit of QCD and 
give a brief presentation of recent progress in the understanding of 
unitarity corrections in this kinematic regime. Special emphasis is given 
to the conformal field theory structure discovered in these unitarity corrections. 
\end{abstract}

\maketitle

\section{Introduction}

To the present day Regge theory \cite{Collins} provides the most successful 
description of hadron--hadron scattering at high energy $s$ 
and small momentum transfer $t$ (see for example \cite{DL}). 
Regge theory is based on 
analyticity and unitarity of the $S$-matrix. The behaviour of 
hadronic scattering amplitudes 
is then encoded in the positions of Regge poles and cuts 
in the plane of complex angular momentum $\omega$. 
Gribov's reggeon calculus \cite{Gribov} provides a consistent field 
theory of interacting reggeons. 

On the other hand, QCD is firmly established as the correct 
microscopic theory of the strong interaction. 
It should therefore be possible to derive Regge theory from 
QCD and thus to understand Regge behaviour 
in terms of quarks and gluons. 
This is one of the most challenging problems in the physics 
of strong interactions and has not been resolved so far. 

The difficulty is caused by the fact that the Regge limit 
is characterized by high parton densities. Moreover, many 
hadronic scattering processes in this kinematics are 
dominated by small momentum scales. 
Both facts indicate that non--perturbative effects 
dominate. 
Fortunately, however, 
there are a few scattering processes that can be treated 
perturbatively even at very high energy, namely those 
involving the scattering of small color dipoles. 
One such process is 
the scattering of highly virtual photons, in which the 
virtuality provides a hard scale. In the following we will 
have this process in mind. Our hope is, of course, that 
we can extract more general features of QCD in the Regge limit 
from our results.

The first step towards  a QCD--based description of 
the high energy limit was done when the 
BFKL Pomeron was derived \cite{BFKL}. 
It describes a $t$-channel exchange with vacuum quantum numbers 
and resums the leading logarithms of the energy $s$ which can 
compensate the smallness of the strong coupling constant. 
We expect the (perturbative) BFKL Pomeron to be applicable 
in processes that are governed by a single hard momentum scale. 
However, the BFKL Pomeron results in a powerlike growth  
$\sigma \sim s^{0.5}$ of the cross--section at high energy. 
This  will eventually lead to a violation of 
the Froissart bound --- a consequence 
of unitarity stating that the growth of hadronic cross--sections can 
at most be logarithmic. In this sense the leading logarithmic 
approximation, i.\,e.\ the BFKL Pomeron, is inconsistent 
and has to be extended in order to restore unitarity. 

\section{BFKL Pomeron and generalized leading logarithmic approximation}

The BFKL Pomeron can be understood as the exchange of a 
bound state of two reggeized gluons in the $t$-channel. 
It is well--known that the restoration 
of unitarity requires the inclusion exchanges 
with more than two reggeized gluons in the $t$-channel, 
leading to the generalized leading logarithmic approximation. 
Therefore we are interested in amplitudes describing the 
production of $n$ (reggeized) gluon in the $t$-channel, 
which can be related to the total cross section. 
These amplitudes 
$D_n^{a_1\dots a_n}(\kf_1, \dots, \kf_n)$ 
depend on the transverse momenta $\kf_i$ of the gluons 
and carry $n$ color labels in the adjoint representation 
of $\mbox{SU}(N_c)$. 
In order to have unitarity in all subchannels the number 
of gluons in the $t$-channel should not be fixed. Therefore 
the amplitudes $D_n$ obey a tower of coupled integral 
equations \cite{Bartels}, the first of which ($n\!=\!2$) 
is identical to the BKFL equation. 
In the Regge limit the dynamics is effectively reduced 
to two dimensions, and consequently these equations 
are, like the BFKL equation, integral equations 
in two--dimensional tranverse momentum space. 
The complex angular momentum $\omega$ 
acts as an energy--like variable in the equations. 
Its conjugate variable, i.\,e.\ rapidity, can thus be 
interpreted as a time--like variable in this context. 

The system of coupled integral equations is very complicated 
and a complete analytic solution seems currently well out of reach. 
Nevertheless, it turns out to be possible to extract very valuable 
information about the structure of the amplitudes. 

\section{Reggeization and field theory structure}

A very important property of QCD is the reggeization 
of the gluon in the Regge limit \cite{Reggeization}. 
This phenomenon manifests 
itself in the emergence of a pole solution of the BFKL 
equation in the color octet channel. The particle corresponding 
to this pole carries the quantum numbers of a gluon: in a certain 
sense, the gluon appears to be a bound state of two gluons. 
This indicates that now the appropriate degrees of freedom 
are reggeized gluons --- collective excitations of the gauge field --- 
rather than elementary gluons. 

A generalization of this phenomenon is found in the higher $n$-gluon 
amplitudes. The equation for the three--gluon amplitude can 
be solved \cite{BW}, 
\begin{eqnarray}
\lefteqn{D_3^{a_1a_2a_3} (\kf_1,\kf_2,\kf_3) = 
C_3 f_{a_1a_2a_3} [ D_2(\kf_1+\kf_2, \kf_3) }
\nonumber \\
&&-D_2(\kf_1+\kf_3, \kf_2) + D_2(\kf_1,\kf_2 +\kf_3)]
\,.
\end{eqnarray}
The three--gluon amplitude is thus a superposition of 
two--gluon amplitudes $D_2$. In each of the three terms %contributions 
two gluons (and in higher amplitudes even more gluons) 
arrange themselves in such a way as to form a 
'more composite' reggeized gluon. (In this process the color part 
of the amplitude is crucial.) As a consequence, an actual 
three--gluon state in the $t$-channel does not 
occur.\footnote{This does not affect the existence of the Odderon, 
since in our case the three--gluon system has even $C$-parity.}

The four--gluon amplitude can be 
shown to have the following structure \cite{BW}: 
\begin{equation}
D_4 = \sum \begin{picture}(0,0)%
\includegraphics{solutiond41.pstex}%
\end{picture}%
\setlength{\unitlength}{3947sp}%
\begingroup\makeatletter\ifx\SetFigFont\undefined
% extract first six characters in \fmtname
\def\x#1#2#3#4#5#6#7\relax{\def\x{#1#2#3#4#5#6}}%
\expandafter\x\fmtname xxxxxx\relax \def\y{splain}%
\ifx\x\y   % LaTeX or SliTeX?
\gdef\SetFigFont#1#2#3{%
  \ifnum #1<17\tiny\else \ifnum #1<20\small\else
  \ifnum #1<24\normalsize\else \ifnum #1<29\large\else
  \ifnum #1<34\Large\else \ifnum #1<41\LARGE\else
     \huge\fi\fi\fi\fi\fi\fi
  \csname #3\endcsname}%
\else
\gdef\SetFigFont#1#2#3{\begingroup
  \count@#1\relax \ifnum 25<\count@\count@25\fi
  \def\x{\endgroup\@setsize\SetFigFont{#2pt}}%
  \expandafter\x
    \csname \romannumeral\the\count@ pt\expandafter\endcsname
    \csname @\romannumeral\the\count@ pt\endcsname
  \csname #3\endcsname}%
\fi
\fi\endgroup
\begin{picture}(865,966)(2373,-2016)
\end{picture}
 + \begin{picture}(0,0)%
\includegraphics{solutiond42.pstex}%
\end{picture}%
\setlength{\unitlength}{3947sp}%
\begingroup\makeatletter\ifx\SetFigFont\undefined
% extract first six characters in \fmtname
\def\x#1#2#3#4#5#6#7\relax{\def\x{#1#2#3#4#5#6}}%
\expandafter\x\fmtname xxxxxx\relax \def\y{splain}%
\ifx\x\y   % LaTeX or SliTeX?
\gdef\SetFigFont#1#2#3{%
  \ifnum #1<17\tiny\else \ifnum #1<20\small\else
  \ifnum #1<24\normalsize\else \ifnum #1<29\large\else
  \ifnum #1<34\Large\else \ifnum #1<41\LARGE\else
     \huge\fi\fi\fi\fi\fi\fi
  \csname #3\endcsname}%
\else
\gdef\SetFigFont#1#2#3{\begingroup
  \count@#1\relax \ifnum 25<\count@\count@25\fi
  \def\x{\endgroup\@setsize\SetFigFont{#2pt}}%
  \expandafter\x
    \csname \romannumeral\the\count@ pt\expandafter\endcsname
    \csname @\romannumeral\the\count@ pt\endcsname
  \csname #3\endcsname}%
\fi
\fi\endgroup
\begin{picture}(865,1333)(3691,-2237)
\put(4427,-1861){\makebox(0,0)[lb]{\smash{\SetFigFont{10}{12.0}{rm}$V_{2 \rightarrow 4}$}}}
\end{picture}
 
\,.
\end{equation}
Here the first (reggeizing) part is again a superposition of two--gluon 
amplitudes, coupled to the photons through a quark box diagram. 
In the second term the $t$-channel evolution starts with a two--gluon 
state that is coupled to a full four--gluon state via a new effective 
2-to-4 transition vertex $V_{2 \rightarrow 4}$. 
The emerging structure is that of a quantum field theory in which 
states with different numbers of gluons are coupled to each other. 
The obvious question is whether this structure found here in the 
four--gluon amplitude persists in higher $n$-gluon amplitudes. 

The necessary analysis of the five-- and six--gluon amplitudes has 
recently been performed \cite{Thesis,d6}. The five--gluon amplitude 
reggeizes completely, i.\,e.\ it is a superposition of two-- and 
four--gluon amplitudes. 
The corresponding mechanism is of more general nature and we expect 
each amplitude with an odd number of gluons to be a superposition 
of lower amplitudes with even numbers of gluons. 

A part of the six--gluon amplitude reggeizes again and can 
be written as a superposition of two-- and four--gluon amplitudes. 
In addition, we have discovered a new piece in the field theory which 
has very interesting symmetry properties. However, further analysis 
is required in order to clarify whether it should be interpreted as 
a two--to--six transition vertex or as a superposition of new 
two--to--four vertices. 
In either case a new element is added to the field theory structure. 

But the most important property of the six--gluon amplitude 
is that it fits nicely into the picture of a field theory structure of 
unitarity corrections. This strengthens our hope that the whole 
set of unitarity corrections can be formulated as an effective 
field theory in $2\!+\!1$ dimensions, with rapidity acting as 
the time--like variable.  

A closer analysis shows that the process of 
reggeization is crucial for the very emergence of the field theory 
structure. It is therefore desirable to further improve our 
understanding of this phenomenon. 
Recent progress in this direction includes the 
formulation of Ward--type identities relating amplitudes 
with different numbers of gluons \cite{Thesis}. 

As a further result the knowledge of the six--gluon amplitude allows 
us to establish 
the existence of a Pomeron--Odderon--Odderon vertex in 
perturbative QCD. 

\section{Conformal invariance}

One of the most interesting properties of the $n$-gluon amplitudes 
is their conformal invariance in impact parameter space. 
After a Fourier transformation from two--dimensional 
transverse momentum space to impact parameter space, 
$\kf_i\! \to \!\vec{\rho}_i$, we can introduce complex coordinates 
$\rho=\rho_x +i \rho_y$. The amplitudes $D_n$ are then 
invariant under M\"obius transformations of the 
coordinates $\rho_i$, 
\begin{equation}
   \rho \rightarrow \rho^\prime = \frac{a \rho +b}{c\rho +d}\,;\:\:\:\:
   ad -bc = 1 
\,.
\end{equation}
This result was found in \cite{Lipatov86} for the BFKL Pomeron, 
and in \cite{BLW} for the 2-to-4 transition vertex 
$V_{2 \rightarrow 4}$. The explicit expression for the new piece 
in the six--gluon amplitude has now been shown to exhibit this 
conformal invariance as well \cite{Thesis,d6}. 
We therefore have reason to hope that the whole effective field 
theory will be conformally invariant in the two dimensions 
of impact parameter space. 

\section{Summary and outlook}

The high energy limit of QCD requires the calculation of 
unitarity corrections to the perturbative Pomeron in the 
generalized leading logarithmic approximation. 
The amplitudes with up to six gluons in the $t$-channel 
have been calculated explicitly. There is strong 
evidence that a beautiful mathematical structure is 
hidden in these unitarity corrections. Specifically, we 
expect that the unitarity corrections can be cast into 
the form of a two--dimensional conformal field theory
in impact parameter space with rapidity as a real parameter. 
This opens the fascinating possibility of applying the powerful 
methods of conformal field theory and ---
once the effective theory is identified --- to derive  
the general features of high energy QCD, now bypassing the 
laborious explicit calculation of higher $n$-gluon amplitudes. 
An even more ambitious aim will eventually be the study of NLO 
corrections in this field theory. 

\section*{Acknowledgements}
I would like to thank J.\ Bartels for fruitful collaboration.


\begin{thebibliography}{9}
\bibitem{Collins}
P.\,D.\,B.\ Collins, 
{\sl An Introduction to Regge Theory and High Energy Physics}, 
Cambridge University Press, Cambridge 1977
\bibitem{DL} P.\,V.\ Landshoff, these proceedings
\bibitem{Gribov} V.\,N.\ Gribov, 
{\sl Sov.\ Phys.\ JETP} {\bf 26}(2) (1968), 414
\bibitem{BFKL} E.\,A.\ Kuraev, L.\,N.\ Lipatov, V.\,S.\ Fadin, 
{\sl Sov.\ Phys.\ JETP} {\bf 45} (1977), 199; 
Ya.\,Ya.\ Balitskii, L.\,N.\ Lipatov, 
{\sl Sov.\ J.\ Nucl.\ Phys.} {\bf 28} (1978), 822
\bibitem{Bartels}
J.\ Bartels, 
{\sl Nucl.\ Phys.} {\bf B 175} (1980), 365; 
DESY 91-074 (unpublished)
\bibitem{Reggeization}
L.\,N.\ Lipatov, 
{\sl Sov.\ J.\ Nucl.\ Phys.} {\bf 23} (1976), 338
\bibitem{BW}
J.\ Bartels, M.\ W\"usthoff, 
{\sl Z.\ Phys.} {\bf C 66} (1995), 157
\bibitem{Thesis}
C.\ Ewerz, Doctoral Thesis, Hamburg University 1998, 
DESY-THESIS-1998-025
\bibitem{d6}
J.\ Bartels, C.\ Ewerz, in preparation
\bibitem{Lipatov86}
L.\,N.\ Lipatov, 
{\sl Sov.\ Phys.\ JETP} {\bf 63}(5) (1986), 904
\bibitem{BLW}
J.\ Bartels, L.\,N.\ Lipatov, M.\ W\"usthoff, 
{\sl Nucl.\ Phys.} {\bf B 464} (1996), 298

\end{thebibliography}
\end{document}